\documentclass[journal]{IEEEtran}
\usepackage{cite}
\usepackage{graphicx}
\usepackage[cmex10]{amsmath}

\begin{document}
\title{Nanoscale Electrodynamic Response of Nb Superconductors}
\author{Tamin Tai$^{*}$, Behnood G. Ghamsari, and Steven M. Anlage% <-this % stops a space

\thanks{Tamin Tai, B. G. Ghamsari and S. M. Anlage are with Center for Nanophysics and Advance Materials (CNAM), Physics Department, University of Maryland, College Park, MD
20742 USA, (* e-mail:tamin@umd.edu)}% <-this % stops a space
\thanks{This work is supported by the US DOE/HEP through grant $\#$ DESC0004950, and also by the ONR AppEl
Center, Task D10, (Award No.\ N000140911190), and CNAM.}}

\maketitle

\begin{abstract}
Our objective is to study the extreme and local electrodynamic
properties of Niobium (Nb), and to relate these properties to specific defects
that limit the ultimate RF performance of superconducting radio
frequency (SRF) cavities made from Nb. Specifically, we wish to
develop a microscopic measure of the response of Nb to RF
magnetic fields up to the critical field ($H_{c}$) at microwave
frequencies (few GHz), and at temperatures down to 4.2 K. In order
to image the local electromagnetic response in the GHz frequency
regime, a magnetic writer from a commercial hard drive is integrated
into the near field microwave microscope and operates with the Nb
sample in the superconducting state. The microwave response of Nb
thin films from a mesoscopic area are found through linear and
nonlinear microwave measurements.
\end{abstract}


\section{Introduction}

\IEEEPARstart{B}{ased} on the needs of the SRF community to identify defects on Nb
surfaces \cite{Muck}, a novel magnetic microscope with the
capability of imaging electrodynamic defects in the high frequency
regime (at least a few GHz) under a strong RF field up to the
thermodynamic critical field of Nb ($\sim$ 200 mT) and at cryogenic
temperatures is desirable. A magnetic writer is an excellent
candidate for creating strong localized field at high frequencies
\cite{ReadRiteCorp}\cite{CMU}. Several studies of the high
frequency characteristics of the magnetic writer shows that it
develops well-confined fields up to a few GHz
\cite{CMU}\cite{Abe}\cite {Koblischka}. An impedance measurement
also shows that the writer is well matched to the microwave source
over a broad frequency range (2 GHz $\sim$ 25 GHz) \cite{Tai}. The
integration of magnetic writers into our group's microwave
microscope also operates successfully at liquid Helium temperature
\cite{Tai}\cite{TaiCPEM}. Therefore the magnetic writer satisfies
our need to measure superconductor electrodynamics in the high
frequency region.
In this paper, we report our experimental results
on Nb thin films with thickness 50 nm. We use the magnetic writer to
induce the linear microwave response and third harmonic signals from
the surface of the Nb thin film, creating response from the linear
Meissner effect, the nonlinear Meissner effect, and vortex nonlinearity.
The nonlinearity from the Nb vortex critical state shows great
potential for high resolution nonlinear imaging to identify defects
on Nb cavity surfaces at accelerator operating frequencies and
temperatures.

\section{Experiment}
\begin{figure}[!t]
\centering
\includegraphics [width=2.5 in]{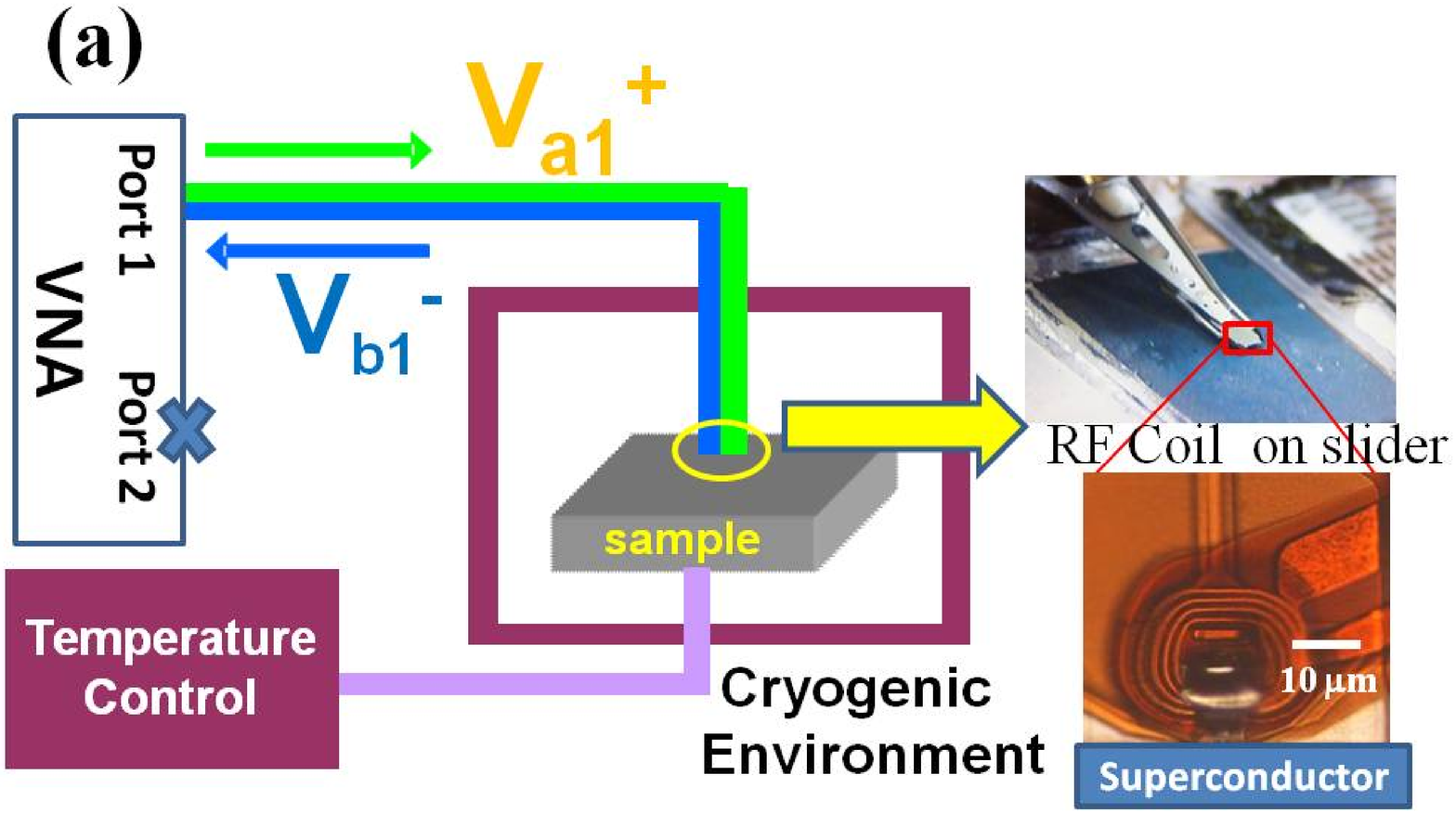}
\includegraphics [width=2.6in, angle=0]{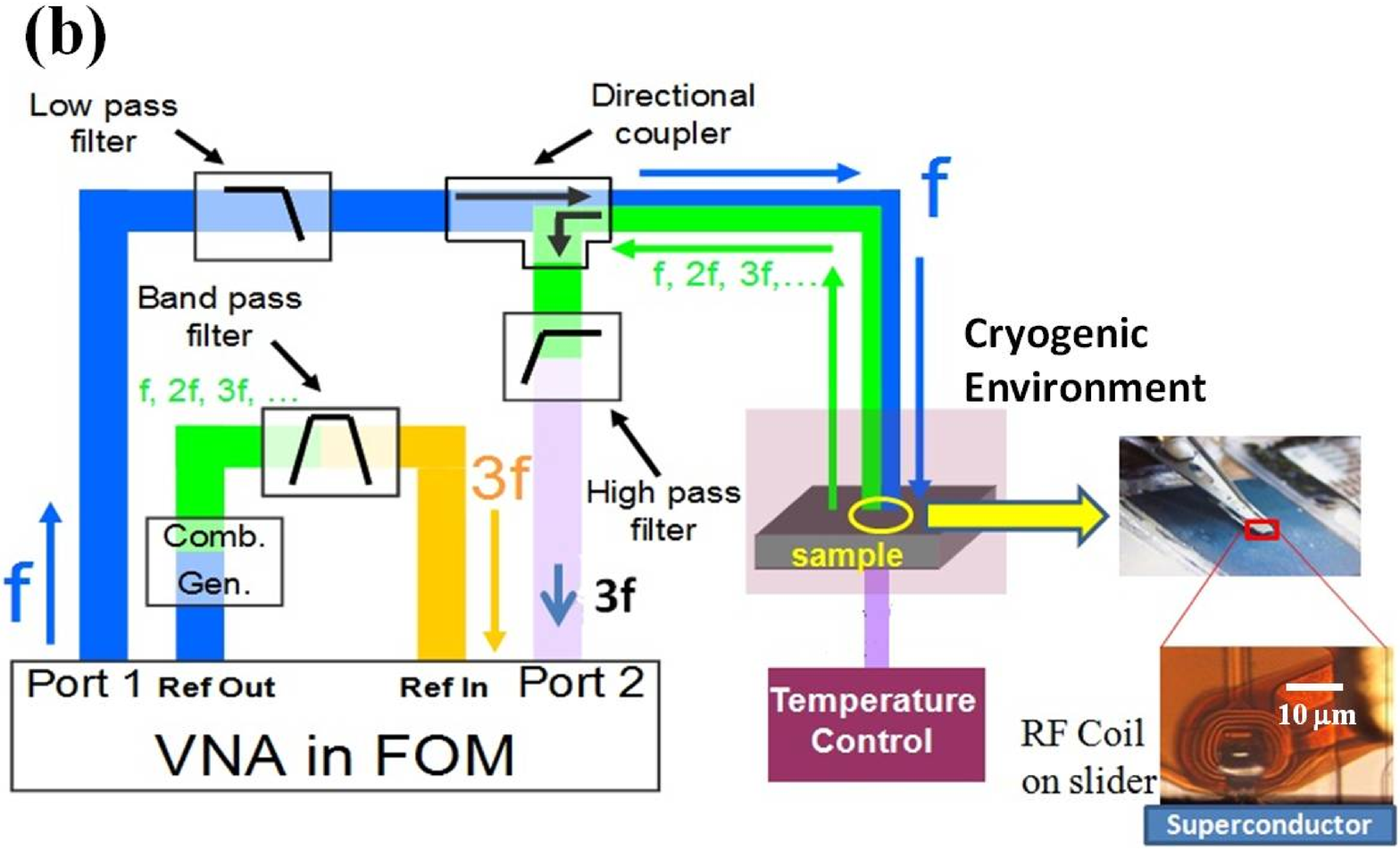}
\caption{(a) Schematic diagram of the linear response measurement, $S_{11}$,
 performed as a function of temperature with the (VNA). (b) Schematic set up of phase-sensitive measurement in nonlinear microwave microscopy. Frequency offset mode (FOM) of the VNA is used in this measurement. The microwave circuit inside the cryogenic environment is the same as (a). Only the circuit outside the cryogenic environment is changed for selectively filtering the $V_{3f}$ signal.}
\label{setup}
\end{figure}
The schematic setup for linear response measurements is shown in Fig.
\ref{setup}(a). A microwave fundamental tone ($V_{a1}^{+}$) is sent
into a Seagate GT5 magnetic writer at one specific frequency from
port 1 of the vector network analyzer (VNA) (model $\sharp$ Agilent
N5242A). The magnetic writer is designed for longitudinal recording
and is integrated into our microwave circuit by soldering the probe
assembly on a coaxial cable. The main part of the writer is a yoke
surrounded by a several turn helical coil which generates the
magnetic flux. The yoke is made of a high permeability material to
channel the magnetic flux to the narrow gap. It is also shielded to
define a nano-scale bit in the recording medium during the writing
process \cite{ReadRiteCorp}-\cite{Koblischka}. Close-up views of the magnetic write head probe on superconducting samples are also shown on the side of Fig. \ref{setup}. In our design, the
magnetic writer approaches the surface of the superconductor in the range of 200 $nm$ $\sim$ 2 $\mu m$. The fundamental tone stimulates
the magnetic writer to generate an RF magnetic field and therefore
excites a screening current on the sample surface so that it can
maintain the Meissner state in the bulk of the material. Larger
magnetic field induces higher screening current within the
penetration depth ($\lambda$) of the superconducting surface, until
the field reaches the critical field of the material. The time
dependent screening current on the superconductor will induce an
electromotive force (emf) on the magnetic writer. The emf voltage
will couple with the incident fundamental tone and reflect back as
an output signal ($V_{b1}^{-}$). We measure the ratio of
$V_{b1}^{-}$ to $V_{a1}^{+}$ ($S_{11}$) at different
temperatures of the superconducting samples. The sample temperature
is controlled by a Lakeshore 340 temperature controller.

The nonlinear amplitude and phase measurements of the superconductor
harmonic response utilizes the two-port VNA method shown in Fig.
\ref{setup}(b). An excited wave (fundamental signal) at frequency $f$
comes from port 1 of the VNA and is low-pass filtered to eliminate
higher harmonics of the source signal. This fundamental tone is sent
to the magnetic write head probe to generate a localized RF magnetic
field on the superconductor sample. The superconducting sample
responds by creating screening currents to maintain the Meissner
state in the material. These currents inevitably produce a
time-dependent variation in the local value of the superfluid
density, and will generate a response at harmonics of the driving
tone. The generated harmonic signals are gathered by the magnetic
probe and returned to room temperature where they are high-pass
filtered to remove the fundamental tone $V_f$. Finally, an
un-ratioed measurement of $V_{3f}$ is performed on port 2 of the
VNA. In order to get a phase-sensitive measurement of the $3^{rd}$
harmonic signal coming from the superconducting sample, a harmonic
generation circuit is connected to provide a reference $3^{rd}$
harmonic signal, and the relative phase difference between the main
circuit and reference circuit is measured. Further detail about this
phase-sensitive measurement technique can be found in Ref. \cite{Mircea}. In this way we measure the complex third harmonic voltage $V_{3f}^{sample}(T)$ or the corresponding scalar power $P_{3f}^{sample}(T)$. The lowest noise floor of the VNA in our
measuring frequency range is -127 dBm for the un-ratioed power
measurement. A ratioed measurement of the complex
$V_{3f}^{sample}(T)/V_{3f}^{ref}$ is also performed at the same
time. An alternative method to lower the noise floor is to remove
the VNA and use a stable synthesizer (model $\sharp$ HP 83620B) on
port 1 and a spectrum analyzer (model $\sharp$: ESA-E E4407B) on
port 2. The noise floor of our spectrum analyzer is -147 dBm. The
microscope thus measures the local harmonic power and phase
generated at the location of the probe, for a given incident
frequency, power and sample temperature.

The superconducting samples we study are Nb thin films with
thickness 50 nm made by sputtering Nb onto 3 inch diameter quartz
wafers. After the deposition, the wafer is diced into many 10*10
$mm^2$ pieces but otherwise left un-disturbed. The Nb sample is well
anchored to the cold plate to ensure that the surface temperature of
the superconductor is the same as the temperature of the cold plate.
The probe is held by a three axis translatable stage. Hence
different points on the surface of the sample can be examined. We
test many pieces from each Nb wafer, and all pieces show consistent
results for their linear and nonlinear microwave response.

\section{Finite Element Simulation}
\begin{figure}
\center
\includegraphics[width=3in]{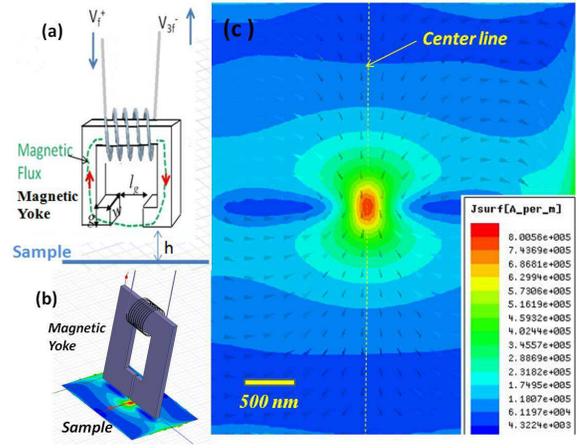}
\caption{(a) A schematic magnetic writer and the gap (not to
scale) with a height (h) away from the sample surface.
In our design, we assume $l_g$=100 nm, $w$=200 nm, and $g$=1 $\mu
m$. (b) Configuration of HFSS simulation of a magnetic writer above a perfectly
conducting sample. (c) Distribution of the surface current density ($J_{surf}$) on the sample surface. The $J_{surf}$ scale bar and arrows indicate the
magnitude and direction of the screening current, respectively, in
the first half of the RF cycle. In this simulation, we assume the
yoke is made of ferrite. The yoke is excited by a 50 $mA$ RF current
and the separation (h) between the probe and the sample is 200 $n
m$.} \label{HFSS}
\end{figure}
\begin{figure}
\centering
\includegraphics[width=2.7in]{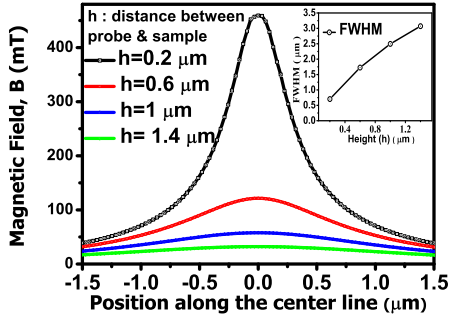}
\caption{Magnetic field (B field) on the sample surface calculated
by HFSS along the horizontal dash center line on the sample in Fig.
\ref{HFSS}. The origin of the sample lies exactly below the center
of the gap and shows the maximum magnetic field. The calculation is
repeated at various heights (h) of the probe above the surface. The inset shows the Full Width Half Maximum (FWHM) of the field distribution as a function of height.}
\label{FieldH}
\end{figure}

The electromagnetic fields produced by the near-field probe on the
superconducting surface can be visualized by a finite element
simulator, ANSYS High Frequency Structure Simulator (HFSS). Fig.
\ref{HFSS} shows the model based on the dimensions of the Seagate
GT5 magnetic writer. In this simulation, the sample is assumed to be
a perfect electric conductor, not a superconductor. The 10 turn coil
is stimulated with a 50 $mA$ drive current at 4.5 GHz excitation
frequency. The magnetic flux induced by the incident waves goes out
through the gap, about 100 $nm$ long ($l_g$), $\sim$ 200 $nm$ wide
($w$), at the end of the yoke. The thickness of the yoke ($g$) is 1
$\mu m$, with a 10 turn gold coil wrapped around it. The sample
develops screening currents with a direction perpendicular to the
direction of magnetic flux on the sample surface. The screening
current on the sample surface has the pattern of a dipole. This
screening response is essentially the same as that developed due to
a horizontal point magnetic dipole at the location of the gap \cite{Peeters}.\\
The surface magnetic field along the center line, shown as a dashed
line in Fig. \ref{HFSS} (c), is plotted in Fig. \ref{FieldH} for
different heights (h) between the probe and the sample. From the
simulation result for a height of 200 $nm$, the maximum magnetic
field (B field) on the surface is 459.2 $mT$, higher than the
thermodynamic critical field of Nb ($\sim$ 200 $mT$). It is clear
that the closer the probe to the sample, the stronger the magnetic
field on the surface. Therefore we want our magnetic writer as close
as possible to the sample. The yoke geometry is
not quite the same as the simple structure schematically shown in
Fig. \ref{HFSS} (a) and (b). Although this simulation cannot
describe the GT5 magnetic writer in detail, it provides a general
trend and prediction for our microscope.

\section{Experimental Results and Discussion}

\begin{figure}
\center
\includegraphics*[width=2.5in]{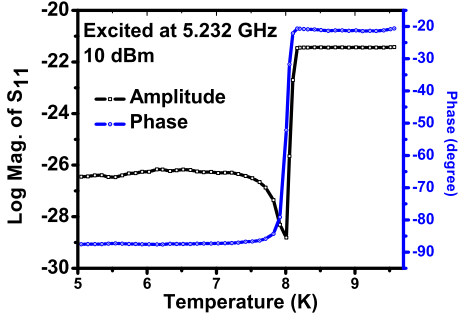}
\caption{The temperature dependent $S_{11}$ of a 50nm Nb thin film
measured with the near-field microwave microscope. Both amplitude
and phase show a transition at 8.3 K, the same temperature as a
global AC susceptibility measurement on this sample.} \label{NbS11}
\end{figure}

Many single-position measurements on different 50 nm thick Nb
samples were performed. Fig. \ref{NbS11} shows the amplitude and
phase of a temperature dependent $S_{11}$ measurement at a
single-position on one of the Nb films under 14 dBm, 5.232 GHz
excitation. A sharp change of $S_{11}$ occurs at 8.3 K for both
amplitude and phase. This change indicates the Nb transition
temperature ($T_c$). Physically, at this temperature, the Nb thin
film becomes superconducting and generates a strong screening
current on the sample surface which couples with the incident
voltage ($V_{a1}^{+}$) and results in a change of the reflected
voltage ($V_{b1}^{-}$). In addition, from the normal state to the
superconducting state, the surface resistance suddenly drops to a
value closer to zero. With a well calibrated measurement and a
suitable circuit model for our system, the penetration depth
and surface resistance of the superconductors can be extracted from
this microwave measurement.

\begin{figure}
\center
\includegraphics[height=2.0in]{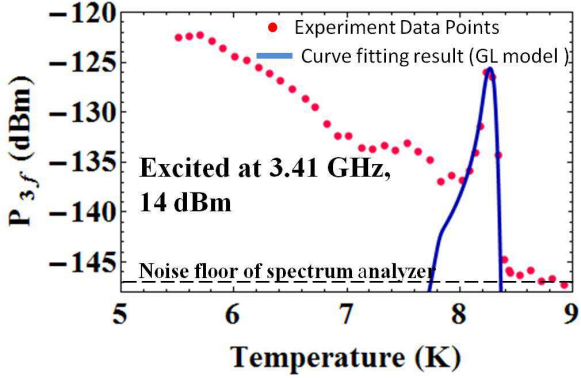}
\caption{Temperature dependence of the third harmonic response for a
50 nm thick Nb film. In this measurement, a microwave synthesizer
and a spectrum analyzer are used for the nonlinear measurement as
port 1 and port 2, respectively on the VNA. the The noise floor of
the spectrum analyzer in the $P_{3f}(T)$ measurement is -147 dBm. The dot points are experiment $P_{3f}$ data. A calculated result (solid curve) based on the GL
model is fit to the data near $T_c$. The parameters are as follows:
$\Gamma$=$10^5$ $A^3/m^2$, $\lambda(0K)$=40 nm,
$\lambda_{cutoff}$=312 nm, $J_{cutoff}$=$2.1*10^{11}$ $A/m^2$, and
$T_c$=8.3 K with a standard deviation of Gaussian spread of $\delta
T_c$=0.03 K.} \label{Nb_P3f_GL}
\end{figure}

Generally speaking, linear response is not very sensitive to the
surface properties. However, for the nonlinear response, the signal
will be very sensitive to the surface defects, for example due to
defects which generate additional channels of dissipation and reactance.
We change our microwave circuit to that in Fig. \ref{setup}(b) and probe many single points on the Nb surface. For each point, the magnetic writer probe has to
lift, laterally translate to another position and then probe the new
position. The height of the probe is judged by optical microscopy
and an attempt is made to keep the same height of the probe in each
measurement. We see significant differences at different points for
this third harmonic nonlinear measurement. A representative result
at one point is shown as the dots in Fig. \ref{Nb_P3f_GL}. A peak in $P_{3f}(T)$
near 8.3 K indicates the $T_{c}$ of this material and can be
interpreted as the intrinsic nonlinearity from the modulation of the superconducting order parameter near $T_c$ due to the decrease of superfluid density and the associated divergence of the penetration depth near $T_{c}$ \cite{Mircea},\cite{SCLee}. Note that a low temperature nonlinearity follows up after the peak and gradually grows while decreasing temperature. This low temperature
nonlinearity may come from the Abrikosov or Josephson vortex critical state, and
implies a nonlinear mechanism related to moving flux or vortices
inside the film \cite{SCLee}. In other words, when the field
from the writer is higher than the lower critical field ($H_{c1}$)
of the Nb, magnetic flux will penetrate into the superconductor and
produce a nonlinear response \cite{Tinkham}. This mechanism can be
interpreted as at least one vortex (or one vortex/antivortex pair)
overcoming the Bean-Livingston barrier \cite{Bean-Livingston} and
penetrating into the film.  In this case, it is not clear how many
vortices are involved or the manner of vortex entry and exit; either
in the form of a semiloop \cite{Gurevich} or perpendicular vortex/
anti-vortex pair \cite{Carneiro}. However, considering the gap
geometry of the magnetic writer probe and the known film thickness,
which is comparable to the value of its magnetic penetration depth
at zero temperature, a harmonic response from a perpendicular
vortex-antivortex pair may be the mechanism of this additional nonlinearity \cite{Carneiro}.\\

The solid line in Fig. \ref{Nb_P3f_GL} shows the result of curve fitting to $P_{3f}$
based on the mechanism of the intrinsic nonlinearity near $T_c$ using the Ginzburg-Landau (GL) model \cite{Lee}. The generated third harmonic power $P_{3f}$ can be estimated as Eq.
(\ref{eq:P3f}) derived for the time dependent nonlinear inductive
circuit \cite{Lee}\cite{LeeThesis}.
\begin{equation}\label{eq:P3f}
    P_{3f}(T)=\frac{\omega^2\mu_{0}^2\lambda^4(T)\Gamma^2(K_{RF})}{32Z_
    {0}d^{6}J^4_{NL}(T)}
\end{equation}
where $\omega$ is the angular frequency of the incident wave,
$\lambda (T)$ is the temperature dependent magnetic penetration
depth, $d$ is the thickness of the film, $\mu_{0}$ is the
permeability of free space, $Z_0$ is the characteristic impedance of
the transmission line in the microscope, and $\Gamma(K_{RF})$ is a
probe geometry factor. The probe geometry factor $\Gamma$ is a
function of the induced surface current density, $K_{RF}$, which is
proportional to the applied magnetic field. The value of geometry
factor is estimated to be $10^5 A^3/m^2$, calculated from the
Karlqvist equation for the magnetic field from a magnetic write gap
\cite{SXWang} under the assumption that the probe height is
around 2 $\mu m$ above the superconducting surface. The term
$J_{NL}$ is the nonlinear current density scale and quantitatively
characterizes the mechanism of nonlinearity. For the intrinsic nonlinearity,
this term is on the scale of the de-pairing current density of the
superconductor. Based on the GL model, the temperature dependent
$J_{NL}(T)$ can be written as
\begin{equation}
J_{NL}=J_0 [1 - (\frac{T}{T_c})^2] [1 -(\frac{T}{T_c})^4]^{1/2}
\end{equation}
where $T_c$ is the critical temperature, $J_{0}$ is the critical
current density at 0 K, which can be estimated as $H_0$/$\lambda_0$
where $H_0$ and $\lambda_0$ are the thermodynamic critical field and
penetration depth of Nb at 0 K, respectively. In our calculation,
$H_0$ is assumed to be 200 mT and $\lambda_0$ is 40 nm. As T
approaches $T_c$, the cutoff of $\lambda (T)$ $\&$ $J_{NL}(T)$ and a
Gaussian distribution of $T_c$ in Eq. (\ref{eq:P3f}) are also
applied in this calculation. Finally from our result of the curve
fitting as shown in the solid curve, this model describes well the
experimental results near $T_c$ and proves that the measured
$P_{3f}(T)$ peak comes from the mechanism of intrinsic nonlinearity. For the
low temperature ($T<T_c$) nonlinearity, other mechanisms must be
responsible for $P_{3f}(T)$.
% new paragraph

\begin{figure}
\center
\includegraphics[height=2.5in]{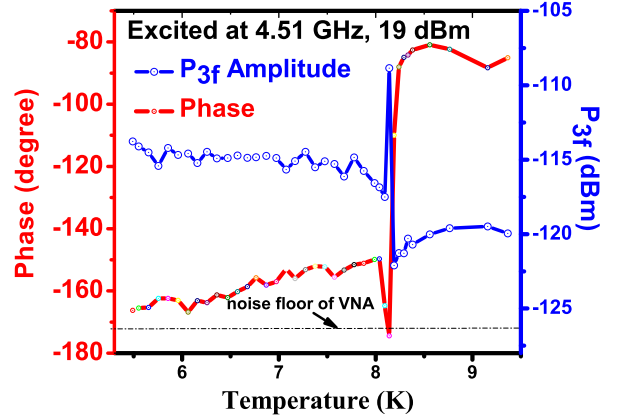}
\caption{Temperature dependence of the third harmonic response with
amplitude (blue) and its relative phase (red) for a 50 nm thick Nb
film. The two-port VNA measurement is used. The noise floor of the
VNA in the $P_{3f}(T)$ measurement is -127 dBm. Note that the temperature scale was corrected because of poor thermal contact of the thermometer.} \label{Nb_P3fPhase}
\end{figure}

In order to further understand the complex-valued third harmonic
voltage $V_{3f}$, the phase-sensitive harmonic technique is
performed on the 50 nm thick Nb thin film. In this measurement, the
unratioed measurement shows a corresponding scalar power
$P_{3f}^{sample}(T)$, which is analogous to the $P_{3f}(T)$
measurement done by the spectrum analyzer. Fig. \ref{Nb_P3fPhase}
shows a representative curve for this measurement. Note that the
probe position in this measurement is different from that of Fig. \ref{Nb_P3f_GL}. The red curve is the phase of the complex
$V_{3f}^{sample}(T)/V_{3f}^{ref}$. The blue curve is the
corresponding scalar power $P_{3f}^{sample}(T)$. Clearly, the phase
shows a minimum at $T_c(\sim 8.3 K)$ while the harmonic magnitude
$P_{3f}(T)$ exhibits a maximum. Note that between the normal state
and the superconducting state of Nb, the phase shift is almost
$\pi/2$, consistent with an analytical model for the harmonic phase
variation near $T_c$\cite{Mircea}. After this minimum, the phase
is followed by a turning point at 8.0 K, which implies a change in
the dominant nonlinear mechanism from the intrinsic nonlinearity near $T_c$ to the
nonlinearity in the vortex critical state. For
temperatures below 8 K, the phase gradually decreases with
decreasing temperatures while the amplitude of the $P_{3f}(T)$
slightly increases.

Comparing to the $P_{3f}(T)$ of Fig. \ref{Nb_P3f_GL}, there are many
similarities and differences. From the $P_{3f}(T)$ above $T_c$
in Fig. \ref{Nb_P3fPhase}, it is clear that the probe shows some
nonlinearity at high excitation power. This nearly temperature independent probe nonlinearity is due to the hysteretic properties of the high permeability material
in the yoke \cite{Tai} and will become significant at high excited
power. Therefore at $T>T_c$, all of the nonlinearity comes from the
probe itself. Both $P_{3f}(T)$ in Fig. \ref{Nb_P3f_GL} and Fig. \ref{Nb_P3fPhase} show a
peak at 8.3 K indicating the $T_c$ of the Nb thin film.
After the peak, both of the temperature dependence of $P_{3f}(T)$ are consistent with that of Josephoson vortices \cite{SCLee}. However, this $P_{3f}(T)$ behaves
a little bit different. We see different levels of $P_{3f}(T)$
increase with decreasing temperatures at different locations. This
may imply that the nonlinearity from the moving vortices is very
sensitive to surface morphology and probe-sample separation. For further detailed analysis, a raster scan of the probe should be performed to realize the relation between surface morphology and vortex motion in the vortex critical state.

\section{Conclusion}
From the linear and nonlinear measurement of Nb
thin films by the near field magnetic field microwave microscope, the electrodynamic properties of Nb materials can be identified. The
linear response can be used to find the film $T_c$ in a local area.
The third harmonic response shows a nonlinearity likely from the
Abrikosov vortex critical state, which implies the magnetic field
from the probe is higher than the $H_{c1}$ of the 50 nm thick Nb
thin film. Therefore the nonlinear near-field magnetic field
microwave microscope has great potential to image the electrodynamic
defects on superconducting Nb in the GHz frequency region.

\end{document}